# Superconductivity at 55 K in iron-based F-doped layered quaternary compound Sm[O$_{1-x}$F$_x$]FeAs

**Ren Zhi-An (任治安)\*, Lu Wei (陆伟), Yang Jie (杨杰), Yi Wei (衣玮), Shen Xiao-Li (慎晓丽), Li Zheng-Cai (李正才), Che Guang-Can (车广灿), Dong Xiao-Li (董晓莉), Sun Li-Ling (孙立玲), Zhou Fang (周放), Zhao Zhong-Xian (赵忠贤)\***

National Laboratory for Superconductivity, Institute of Physics and Beijing National Laboratory for Condensed Matter Physics, Chinese Academy of Sciences, P. O. Box 603, Beijing 100190, P. R. China

**Abstract:**

Here we report the superconductivity in the iron-based oxyarsenide Sm[O$_{1-x}$F$_x$]FeAs, with the onset resistivity transition temperature at 55.0 K and Meissner transition at 54.6 K. This compound has the same crystal structure as LaOFeAs with shrunk crystal lattices, and becomes the superconductor with the highest critical temperature among all materials besides copper oxides up to now.

The equiatomic transition metal quaternary oxypicnides have been studied for long years [1-2], some Fe- and Ni-based oxypicnides have been found to be superconducting at low temperatures recently [3-4], and the very recent discovered F-doped La[$O_{1-x}F_x$]FeAs with a superconducting critical temperature ($T_c$) of 26 K [5], is of great interest because of its higher $T_c$, layered structure and iron-containing character. The later experiments with the replacement of La by other rare earth elements, such as Sm, Ce, Pr, Nd *etc.* [6-9], has put this class to another high-$T_c$ family that superconducts above 50 K. All these arsenide (including phosphide) superconductors formed in a same tetragonal layered structure with the space group P4/nmm that has an alternative stacked Fe-As layer and Re-O (Re = rare earth metals) layer, and the $T_c$ is observed to be increased by the smaller rare earth substitution with shrunk crystal lattice. Here we report our new results on the samarium-arsenide Sm[$O_{1-x}F_x$]FeAs that synthesized under high pressure, with a resistivity onset $T_c$ of 55.0 K, which is the highest among all materials besides copper oxides up to now.

The superconducting Sm[$O_{1-x}F_x$]FeAs samples were prepared by a high pressure (HP) synthesis method [9]. SmAs powder (pre-sintered) and As, Fe, $Fe_2O_3$, $FeF_2$ powders (the purities of all starting chemicals are better than 99.99%) were mixed together according to the nominal stoichiometric ratio of Sm[$O_{1-x}F_x$]FeAs, then ground thoroughly and pressed into small pellets. The pellets were sealed in boron nitride crucibles and sintered in a high pressure synthesis apparatus under the pressure of 6 GPa and temperature of 1250°C for 2 hours. Comparing with the common vacuum quartz tube synthesis method, the HP method is more convenient and efficient for synthesize gas-releasing compound with super-high pressure-seal. The structure of the samples was characterized by powder X-ray diffraction (XRD) analysis on an MXP18A-HF type diffractometer with Cu-$K_\alpha$ radiation from 20° to 80° with a step of 0.01°.

The XRD patterns indicate that all samples have a main phase of LaOFeAs structure with some impurity phases, and the impurity phases have been determined to be the known oxides, arsenides, and fluorides that were formed by starting chemicals, which do not superconduct at the measuring temperature. Here we note that because of the inevitable loss of fluorine either by HP synthesis or ambient pressure synthesis, the real F-doped level is much smaller than the nominal one, and therefore the impurity phases always exist due to the unbalance of the stoichiometry for the nominal phase. The lattice parameters for all samples were calculated by the least-square fit method with $|\delta 2\theta| < 0.01°$. For the undoped SmOFeAs, the lattice parameters a = 3.933(5) Å, c = 8.495(4) Å, while all superconducting samples have smaller lattices; for the nominal Sm[$O_{0.9}F_{0.1}$]FeAs, a = 3.915(4) Å and c = 8.428(7) Å. This result is different with the previous reported data where the crystal lattice was enlarged by F-doping [6] while consistent with all reports of other rare earth substitutions, and indicates the covalent character of the intra-layer chemical bonding due to the smaller covalent radius of fluorine than oxygen.

The DC resistivity was measured by the standard four-probe method. The results for the HP sample with a nominal composition of Sm[$O_{0.9}F_{0.1}$]FeAs and an undoped SmOFeAs sample (sintered in sealed vacuum quartz tube) are shown in Fig. 2. The resistivity of SmOFeAs shows an anomaly at 150 K, which is similar to that of other ReOFeAs compounds that reported previously [5, 7], and this anomaly was confirmed to be caused by the occurrence of spin density wave instability [12]. For the Sm[$O_{0.9}F_{0.1}$]FeAs, the temperature of the onset resistivity transition was found to be at 55.0 K and the zero resistivity appeared at 52.6 K, which is higher than that of Pr[$O_{1-x}F_x$]FeAs and Nd[$O_{1-x}F_x$]FeAs, and then becomes *the highest* among all superconducting materials besides copper oxides. As Sm has a smaller covalent radius comparing with La, Ce, Pr and Nd, the inner chemical pressure that caused by the shrinkage of crystal lattice is thought to be an important factor that enhances $T_c$ [10], as proposed in a theoretical calculation in Ref. [11], where it is indicated that the $T_c$ may be enhanced by the increase of hopping integral, which can be achieved by the shrinkage of the lattice.

The magnetization measurements were performed on a Quantum Design MPMS XL-1 system during warming cycle under fixed magnetic field after zero field cooling (ZFC) and field cooling (FC) process. The DC-susceptibility data (measured under a magnetic field of 1 Oe) are shown in Fig. 3. The sharp magnetic transitions on the DC susceptibility curves indicate the good quality of this superconducting component. The onset diamagnetic transition starts at 54.6 K, and the 10% and 90% transitions on the ZFC curve are at 52 K and 50.6 K respectively, with the middle of this Meissner transition at 51.5 K. For this class with much smaller rare earth substitution, higher $T_c$ might be expected but samples with clear phase of the same structure is still absent.

Acknowledgements:

We thank Mrs. Shun-Lian Jia for her kind helps in resistivity measurements. This work is supported by Natural Science Foundation of China (NSFC, No. 10734120 and No. 50571111) and 973 program of China (No. 2006CB601001 and No. 2007CB925002). We also acknowledge the support from EC under the project COMEPHS TTC.

Corresponding Author:

Zhi-An Ren: renzhian@aphy.iphy.ac.cn

Zhong-Xian Zhao: zhxzhao@aphy.iphy.ac.cn

References:

[1]. Zhu W. J., Huang Y. Z., Dong C., Zhao Z. X., *Materials Research Bulletin*, **29** (1994) 143.

[2]. Zimmer B. I., Jeitschko W., Albering J. H., Glaum R., Reehuis M., *J. Alloys Compd*. **229** (1995) 238.

[3]. Kamihara Y., Hiramatsu H., Hirano M., Kawamura R., Yanagi H., Kamiya T., Hosono H., *J. Am. Chem. Soc*. **128** (2006) 10012.

[4]. Watanabe T., Yanagi H., Kamiya T., Kamihara Y., Hiramatsu H., Hirano M., Hosono H., *Inorg. Chem*. **46** (2007) 7719.

[5]. Kamihara Y., Watanabe T., Hirano M. and Hosono H., *J. Am. Chem. Soc.* **130** (2008) 3296.

[6]. Chen X. H., Wu T., Wu G., Liu R. H., Chen H. and Fang D. F., *Condmat:arXiv*, 0803-3603 (2008).

[7]. Chen G. F., Li Z., Wu D., Li G., Hu W. Z., Dong J., Zheng P., Luo J. L. and Wang, N. L., *Condmat:arXiv*, 0803- 3790 (2008).

[8]. Ren Z. A., Yang J., Lu W., Yi W., Che G. C., Dong X. L., Sun L. L. and Zhao Z. X., *Cond-mat:arXiv*, 0803.4283 (2008).

[9]. Ren Z. A., Yang J., Lu W., Yi W., Shen X. L., Li Z. C., Che G. C., Dong X. L., Sun L. L., Zhou F. and Zhao Z. X., *Cond-mat:arXiv*, 0803.4234 (2008).

[10]. Lu W., Yang J., Dong X. L., Ren Z. A., Che G. C. and Zhao Z. X., *Cond-mat:arXiv*, 0803-4266 (2008).

[11]. Han Q., Chen Y. and Wang Z. D., *Cond-mat:arXiv*, 0803- 4346 (2008).

[12]. McGuire M. A., Christianson A. D., Sefat A. S.,Jin R., Payzant E. A., Sales B. C., Lumsden M. D., Mandrus D., *Cond-mat:arXiv*, 0803- 0796 (2008).

Figure captions:

Figure 1: X-ray powder diffraction pattern of the nominal Sm[$O_{0.9}F_{0.1}$]FeAs compound; the vertical bars correspond to the calculated diffraction intensities.

Figure 2: The temperature dependence of resistivity for the undoped SmOFeAs and the Sm[$O_{0.9}F_{0.1}$]FeAs superconductor.

Figure 3: The temperature dependence of the DC-susceptibility and differential ZFC curve for the Sm[$O_{0.9}F_{0.1}$]FeAs superconductor.

Figure 1:

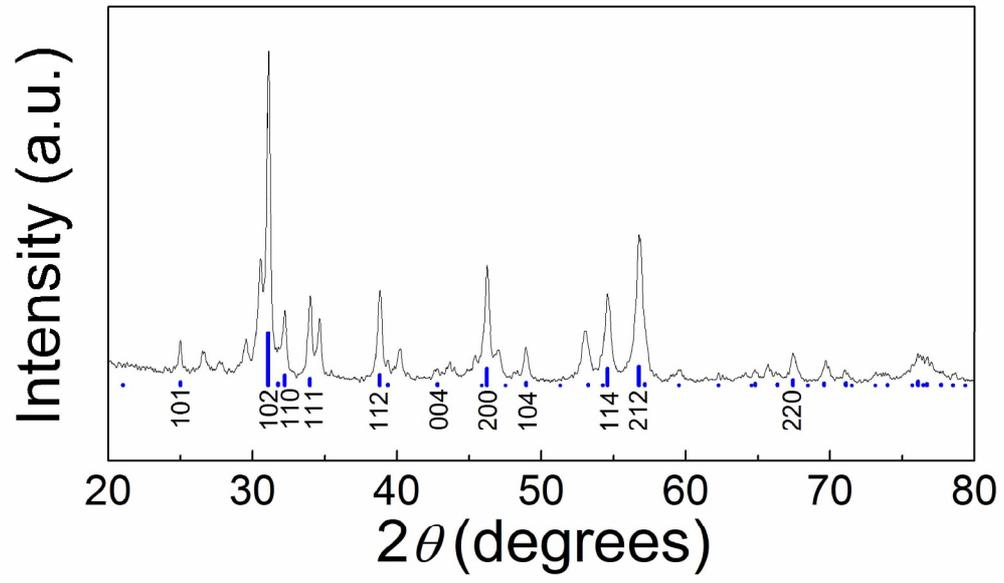

Figure 2:

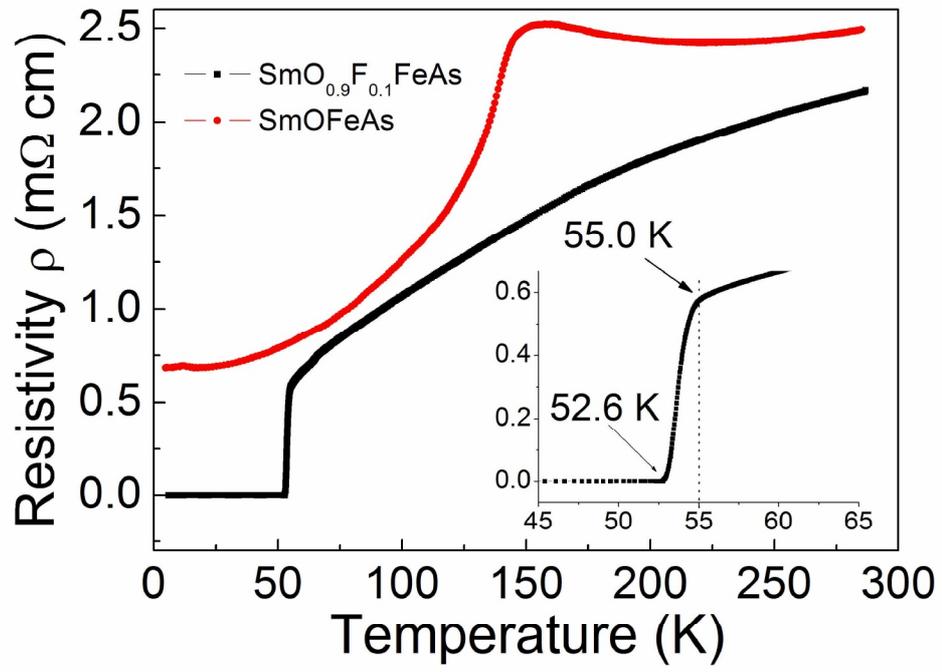

Figure 3:

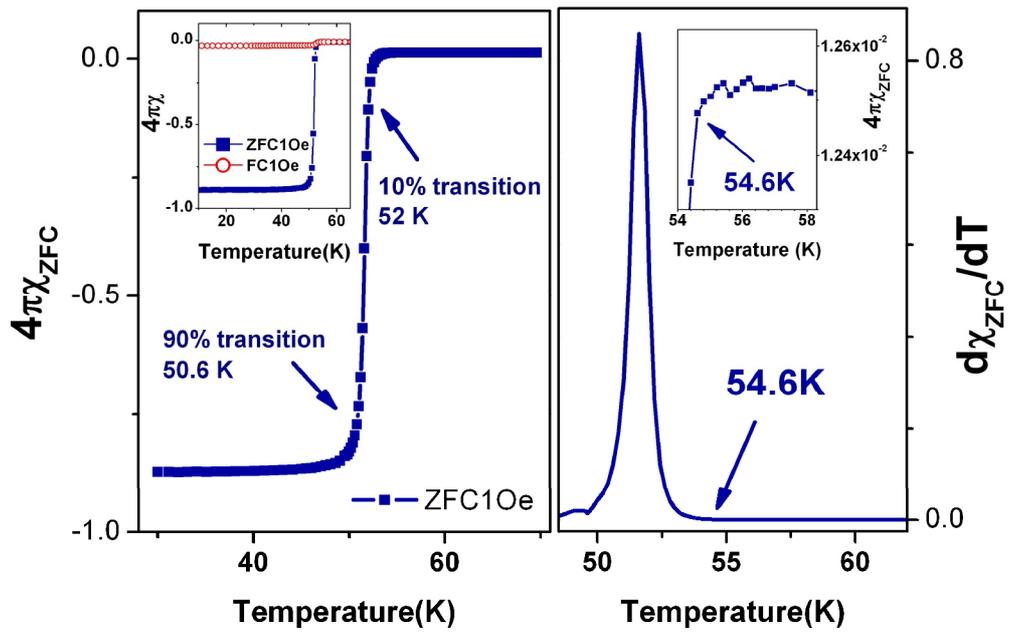